\documentclass{article}
\usepackage{arxiv}

\usepackage[utf8]{inputenc}
\usepackage[T1]{fontenc}
\usepackage{physics}
\usepackage{siunitx}
\usepackage{hyperref}       
\usepackage{url}            
\usepackage{booktabs}       
\usepackage{amsfonts}       
\usepackage{nicefrac}       
\usepackage{microtype}      
\usepackage{graphicx}
\usepackage{doi}

\newcommand*{\figref}[2][]{%
  \hyperref[{#2}]{%
    \ref*{#2}%
    \ifx\\#1\\%
    \else
      #1%
    \fi
  }%
}

\title{EigenCWD: a spatially-varying deconvolution algorithm for single metalens imaging}

\date{} 					

\author{%
\href{https://orcid.org/0000-0001-5160-7628}{\includegraphics[scale=0.06]{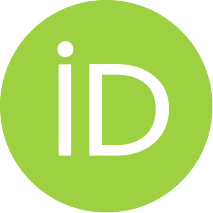}\hspace{1mm}Joel Yeo$^{1,2,4}$}
\quad
\href{https://orcid.org/0000-0002-8886-510X}{\includegraphics[scale=0.06]{orcid.pdf}\hspace{1mm}\textbf{N. Duane Loh}$^{2,3}$} 
\quad
\href{https://orcid.org/0000-0001-7836-681X}{\includegraphics[scale=0.06]{orcid.pdf}\hspace{1mm}Ramon Paniagua-Dominguez$^{4,+}$} 
\quad
\href{https://orcid.org/0000-0002-7622-8939}{\includegraphics[scale=0.06]{orcid.pdf}\hspace{1mm}Arseniy I. Kuznetsov$^{4,*}$} 
\\
$^1$NUS Graduate School for Integrative Sciences and Engineering Programme, National University of\\Singapore, 119077 Singapore, Singapore, \\ 
$^2$Department of Physics, National University of Singapore, 117551 Singapore, Singapore,\\
$^3$Department of Biological Sciences, National University of Singapore, 117558 Singapore, Singapore,\\
$^4$Institute of Materials Research and Engineering (IMRE), Agency for Science, Technology and
Research\\(A*STAR), 2 Fusionopolis Way, Innovis \#08-03, 138634 Singapore, Singapore,\\
\\
$^+$\texttt{Corresponding author: Ramon\_Paniagua@imre.a-star.edu.sg}
\\
$^*$\texttt{Corresponding author: Arseniy\_Kuznetsov@imre.a-star.edu.sg}
}

\renewcommand{\shorttitle}{Ghostbuster: a phase retrieval diffraction tomography algorithm for cryo-EM}

\hypersetup{
pdftitle={Ghostbuster: a phase retrieval diffraction tomography algorithm for cryo-EM},
pdfsubject={},
pdfauthor={Joel Yeo, Benedikt J. Daurer, Dari Kimanius, Deepan Balakrishnan, Tristan Bepler, Yong Zi Tan, N. Duane Loh},
}

\begin{document}

\maketitle




\begin{abstract}
The miniaturization of optics through the use of two-dimensional metalenses has enabled novel applications in imaging.
To date, single-lens imaging remains the most common configuration, in part due to the limited focusing efficiency of metalenses.
This results in limitations when it comes to wavefront manipulation and, thus, unavoidable aberrations in the formed image that require computational deconvolution to deblur the image.
For certain lens profiles, such as the most common hyperbolic one that results in the highest efficiencies, at large fields of view, spatially-varying aberrations such as coma or astigmatism are prominent.
These aberrations cannot be corrected for by traditional deconvolution methods, such as Wiener filtering.
Here, we develop a spatially-varying deconvolution algorithm based on eigenvalue column-wise decomposition (eigenCWD).
EigenCWD solves a minimization problem of the error between the measured image and the estimated image of the object to be reconstructed through an approximate forward blurring model.
This approximate forward model uses an eigendecomposition of the spatially-varying point spread functions for fast computation, allowing for efficient scaling to larger image sizes and blurring kernels common in metalens imaging.
We demonstrate eigenCWD's ability to correct spatially-varying blur and distortions for various lens profiles, surpassing that of the Wiener filter.
\end{abstract}

\section{Introduction}

Metalenses are a particular application of metasurfaces designed specifically for collecting and/or focusing light, which is widely used in imaging, sensing, metrology, and lighting \cite{Kuznetsov2024-np}.
The ability to replicate any phase profile on a metalens whilst maintaining an ultra-thin form factor presents greater versatility over conventional bulky lenses \cite{Kuznetsov2016-gy, Genevet2017-cd, Pan2022-uk}.
This has been demonstrated in various imaging applications such as microscopy, endoscopy, multi and hyperspectral imaging, and even astronomy.
In general, however, the moderate focusing efficiency of many metalenses restricts their use to single-lens imaging, resulting in unavoidable aberrations and geometric distortions in the image formed.
As such, it is typical to complement metalens imaging with computational post-processing methods such as deconvolution to remove aberrations \cite{Peng2016-ae, Colburn2018-ww, Colburn2020-ak, Huang2020-tn, Zhao2021-kr}.
Traditional deconvolution approaches include Wiener filtering \cite{Wiener2019-je}, and iterative algorithms such as the Richardson-Lucy \cite{Richardson1972-rs, Lucy1974-rw} and total-variation (TV) regularization-based \cite{Acar1994-df} deconvolution.

These traditional algorithms assume a spatially invariant blur kernel, and therefore cannot correct for geometric distortions or spatially-varying aberrations such as coma.
Alternatively, deblurring metalens images through deep learning approaches has been gaining traction over the years \cite{Dun2020-uh, Tseng2021-ei, Tan2021-dk, Fan2022-dg, Hu2023-ys, Pinilla2023-ll, Maman2023-ci, Liu2024-fe, Cheng2024-ie, Zhang2024-gu}.
Using a large dataset with both ground truth images and their blurred counterparts, deep networks can statistically learn to deblur a blurred metalens image.
However, these deep learning deblurring approaches require long hours of training and curating datasets for a specific lens profile and therefore are not readily generalizable for a wide range of lenses.

Accounting for spatially-varying aberrations first requires an appropriate forward model for the spatially-varying blur.
This can be written as a linear operator $\vb{H}\in \mathbb{R}^{M^2 \times M^2}$, where we assume the object is a square of size $M\times M$ for simplicity.
This polynomial scaling in memory for larger image sizes makes it impractical to solve the inverse problem of deblurring directly.
As such, the wide range of spatially-varying deblurring algorithms developed by many over the past few decades involve strategies to reduce the memory requirements for evaluating the forward blurring model.
These may be broadly categorized into two families.
The first of them assumes local point spread function (PSF) invariance within sub-regions of the object, and therefore performs patch-wise deconvolution followed by appropriate stitching methods to recover the full object \cite{Nagy1997-xc, Hirsch2010-mx, Denis2015-ol, Bardsley2006-zj}.
This approach may be prone to border artifacts in the reconstructed object at the edges between sub-regions, which may be mitigated through appropriate windowing functions or using overlapping sub-regions.
The second family of algorithms decomposes the spatially-varying PSFs such that they may be expressed as a weighted sum of invariant basis PSFs \cite{Lauer2002-au, Sroubek2016-ki, Turcotte2020-gl, Novak2021-pv}, allowing for efficient computation using the convolution theorem.
The smooth interpolation for calculating the weights (more details in Sec.~\ref{sec:eigenpsf}) removes any border artifacts in the forward model.
Furthermore, by choosing an appropriate decompositional basis, algorithmic speedups are possible by truncating the least significant components. 

Here, we present a spatially-varying deconvolution algorithm based on eigenvalue column-wise decomposition (eigenCWD) of the linear operator $\vb{H}$, whose columns are the flattened spatially varying PSFs.
EigenCWD is a modification of the column-wise decomposition via singular value decomposition (CWD-SVD) algorithm developed by Sroubek \cite{Sroubek2016-ki}, where we use the eigenPSF decomposition method described by Lauer \cite{Lauer2002-au} for the forward blurring model.
This circumvents the need to construct the large $\vb{H}$ matrix in Sroubek's original implementation, allowing for practical scaling to much larger images and blur kernels suitable for metalens imaging.
Spatially-varying deconvolution is then performed by solving the inverse problem using the alternating direction method of multipliers (ADMM) method \cite{Boyd2011-mm} with a total variation (TV) regularizer.
With this approach, we demonstrate significant improvements with respect to these methods in computational cost and in reconstruction quality compared to the more standard Wiener filtering for various metalens profiles.

\section{EigenPSF: decomposition of spatially-varying point spread functions}
\label{sec:eigenpsf}

The development of a spatially-varying deconvolution algorithm first requires an efficient method of computing the forward model to simulate spatially-varying blur.

Image formation by an incoherent imaging system that can be characterized by an invariant PSF, $p_\mathrm{inv}$, can be modeled as a convolution:
\begin{align}\label{eq:convolution}
    g_\mathrm{inv}\qty(x,y) = \qty[f\otimes p_\mathrm{inv}]\qty(x,y).
\end{align}
where $\qty(x,y)$ denotes the spatial coordinates of the image plane, $f$ is the object, $g_\mathrm{inv}$ is the image, and $\otimes$ is the convolution operator.
The convolution in Eq.~\eqref{eq:convolution} can be efficiently computed using the convolution theorem for this simple model.

In general, however, the PSF of any incoherent imaging system varies as a function of the spatial positions of the point emitters.
Therefore, the generalized image formation is instead described by
\begin{align}\label{eq:rs}
    g\qty(x,y) = \iint f\qty(u,v)p\qty(u,v,x-u,y-v)\dd{u}\dd{v},
\end{align}
where $\qty(u,v)$ denotes the spatial coordinates of the object plane, and $p$ is the corresponding PSF at this point.
In this form, the convolution theorem no longer applies to Eq.~\eqref{eq:rs}.
This therefore requires explicit calculation of the integrals, which is computationally inefficient for most deconvolution algorithms.

To overcome this, a common strategy is to decompose the spatially-varying PSFs into a weighted, linear sum of basis PSFs \cite{Lauer2002-au, Novak2021-pv, Turcotte2020-gl,  Sroubek2016-ki}:
\begin{align}\label{eq:decompose}
    p\qty(u,v,x,y) = \sum_{i=1}^\infty a_i\qty(u,v) q_i\qty(x,y),
\end{align}
where $q_i$ are the basis PSFs and $a_i$ are the corresponding weights that encode the variation of the PSFs over the finite extent of the object.
Substituting Eq.~\eqref{eq:decompose} into Eq.~\eqref{eq:rs}, we can rewrite the generalized image formation model as
\begin{align}\label{eq:eigenconvolution}
    g\qty(x,y) &= \sum_{i=1}^\infty\iint f\qty(u,v) a_i\qty(u,v) q_i\qty(x-u,y-v)\dd{u}\dd{v} = \sum_{i=1}^\infty \qty[\qty(fa_i) \otimes q_i]\qty(x,y).
\end{align}
Eq.~\eqref{eq:eigenconvolution} describes a sum of convolutions, which can now be computed much more efficiently using the convolution theorem as opposed to Eq.~\eqref{eq:rs}.

\begin{figure}[ht]
    \centering
    \includegraphics[width=0.5\textwidth]{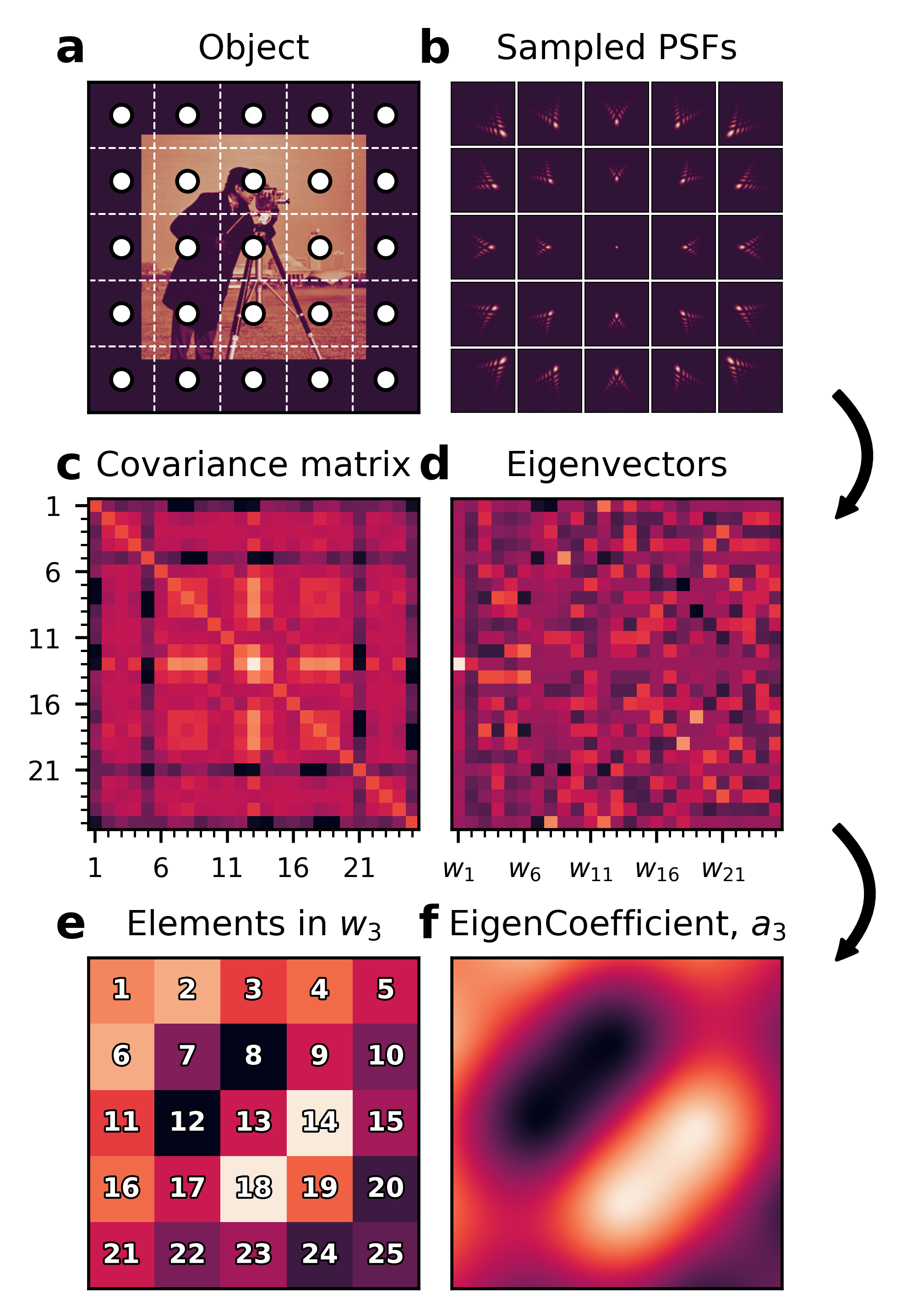}
    \caption{Schematic of the eigenPSF formalism for 25 PSFs sampled on a $5 \times 5$ grid. 
    (a) The white circles denote the $5\times 5$ spatial locations of point sources which give rise to (b) the corresponding 25 PSFs, $p_i$. 
    (c) A $25 \times 25$ covariance matrix, $\vb{C}$, is computed from these PSFs, (d) whose eigenvectors, $\qty{\vb{w}_1, \vb{w}_2, \ldots, \vb{w}_{25}}$, are depicted here as columns. 
    (e) Using the 3rd eigenvector, $\vb{w}_3$ as an example, its elements (labeled by the numbers in each pixel), denote the weight value (color) at the original spatial locations of the point sources. 
    (f) Interpolation is performed to obtain the weights at every pixel in the object coordinates, which gives rise to the eigencoefficient, $a_3$.}
    \label{fig:eigenprocess}
\end{figure}

The weights and basis PSFs can be computed by performing a linear decomposition on a set of $N$ PSFs, $\qty{p_1,p_2,\ldots,p_N}$, sampled from different spatial locations in the object.
Common decomposition methods include eigendecomposition (also known as principle component analysis \cite{Baena-Galle2015-fd, Irace2016-yq} or Karhunen-Loeve decomposition \cite{Lauer2002-au, Novak2021-pv}), and singular value decomposition \cite{Sroubek2016-ki} (SVD).
Here, we use the eigendecomposition approach (see Fig.~\ref{fig:eigenprocess}) which consists of the following steps:
\begin{enumerate}
    \item Compute the covariance matrix, $\vb{C}$, for the set of sampled PSFs, $\qty{p_1,p_2,\ldots,p_N}$:
    \begin{align}
        \vb{C}_{ij} = \mathrm{cov}\qty[p_i, p_j] \qq{where} i,j=1,2,\ldots,N.
    \end{align}

    \item Calculate the eigenvalues, $\lambda_i$, and eigenvectors, $\vb{w}_i$, of $\vb{C}$ by solving:
    \begin{align}
        \vb{C}\vb{w}_i = \lambda_i\vb{w}_i.
    \end{align}
    The eigenvalues $\qty{\lambda_1, \lambda_2, \ldots, \lambda_N}$ are to be sorted in descending order of magnitude.

    \item The eigenPSFs are therefore constructed as
    \begin{align}
        q_i = \sum_{j=1}^N \qty(\vb{w}_i)_jp_j,
    \end{align}
    where $\qty(\vb{w}_i)_j$ denotes the $j$th element of vector $\vb{w}_i$.

    \item The weights to decompose the $j$th sampled PSF into the eigenPSF basis are given by the $j$th elements in the $\vb{w}_i$ eigenvectors:
    \begin{align}\label{eq:sumeigenpsf}
        p_j = \sum_{i=1}^N \qty(\vb{w}_i)_jq_i \approx \sum_{i=1}^K \qty(\vb{w}_i)_jq_i,
    \end{align}
    where choosing $K<N$ allows for faster computation by truncating the number of components to use.
    This assigns the elements in the $\vb{w}_i$ eigenvectors to the spatial locations from which the original PSFs were sampled.
    Interpolation is then used to calculate the weights for all other points of the object to obtain the eigencoefficients, $a_i\qty(u,v)$, in Eq.~\eqref{eq:eigenconvolution}.
\end{enumerate}

\section{EigenCWD}
\label{sec:eigencwd}

\begin{figure}[ht]
    \centering
    \includegraphics[width=\textwidth]{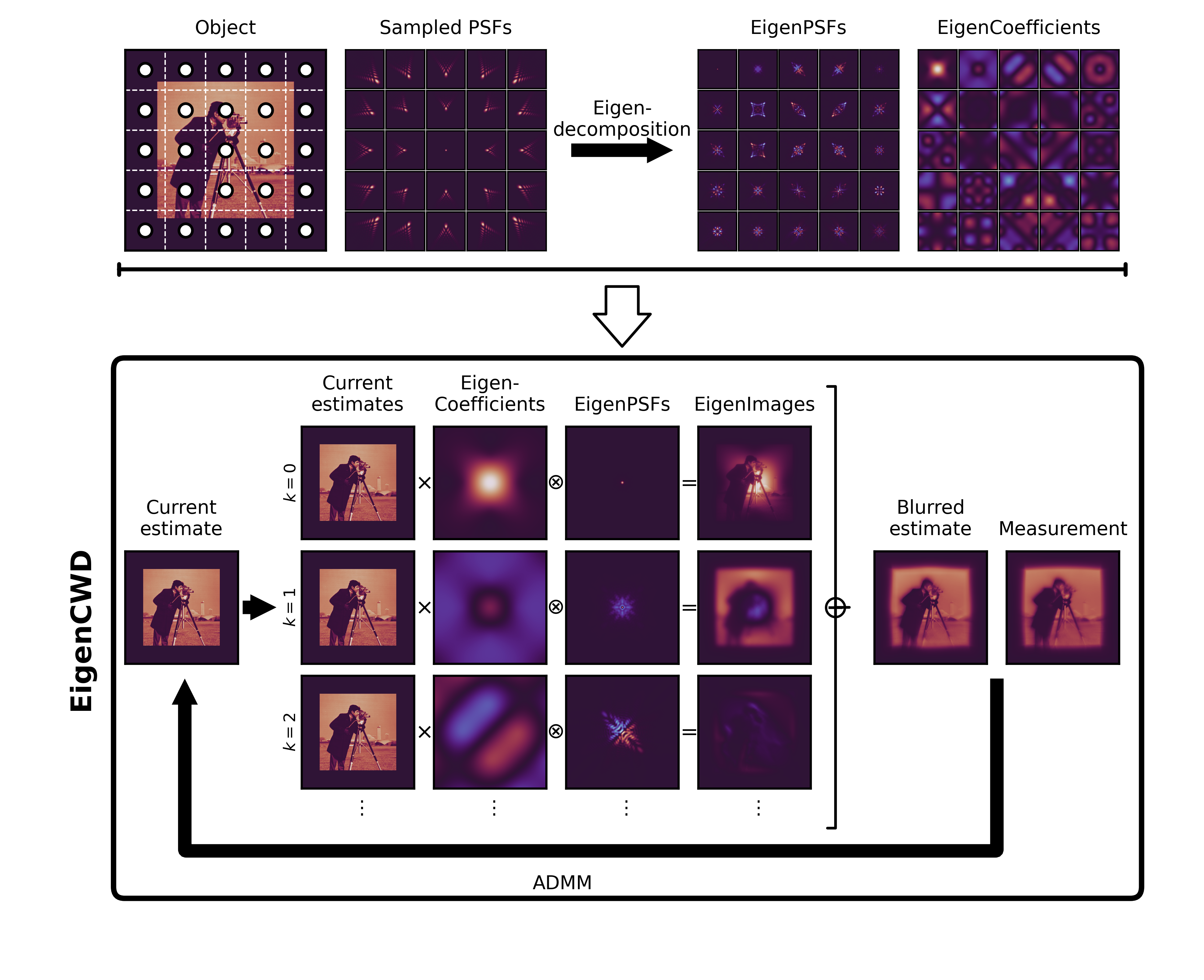}
    \caption{Schematic of the eigenCWD algorithm. The object is blurred using the eigenPSF forward model, and deblurred via solving the inverse problem with eigenCWD.}
    \label{fig:eigencwd_schematic}
\end{figure}

The eigenPSF formalism outlined in the previous section represents an efficient approach to compute the forward model of an object with spatially-varying blur.
In this section, we present our eigenCWD algorithm (Fig.~\ref{fig:eigencwd_schematic}) which solves the inverse problem of spatially deblurring the image based on the eigenPSF formalism.
EigenCWD is based on the CWD-SVD algorithm developed by Sroubek \cite{Sroubek2016-ki}, but modified to avoid the computation of unnecessary large matrices which makes it scale poorly with the sizes of objects and PSFs typical in metalens imaging.

To describe the eigenCWD algorithm, we first rewrite Eq.~\eqref{eq:rs} in matrix notation, assuming an object and image of size $M \times M$:
\begin{align}\label{eq:Hf}
    \vb{g} = \vb{Hf},
\end{align}
where $\vb{g}, \vb{f} \in \mathbb{R}^{M^2}$ are the object and image arrays flattened into a 1D column vector, and
\begin{align}\label{eq:matrix}
    \vb{H} = \begin{pmatrix}
        \vb{p}_1 & \mid & \vdots & &  \\
        \mid & \vb{p}_{2} & \mid &  &  \\
        \vdots & \mid & \vb{p}_{3} & & \vdots \\
         & \vdots & \mid & \ddots & \mid \\
         &  & \vdots & & \vb{p}_{M \times M} \\
    \end{pmatrix},
\end{align}
where $\vb{H} \in \mathbb{R}^{M^2 \times M^2}$ is the transfer matrix which contains the spatially-varying PSFs, $\vb{p}_i$, in the $i$th column which corresponds to the $i$th pixel location in the object.
Performing the matrix product in Eq.~\eqref{eq:Hf} is equivalent to computing the full double integral for the generalized image formation in Eq.~\eqref{eq:rs}.
For large $M$, it becomes highly inefficient to compute the PSF for each of the $M^2$ pixels, and also memory intensive to perform the matrix multiplication in Eq.~\eqref{eq:matrix}.

The eigenPSF formalism described in the previous section removes the need to determine $\vb{H}$ by decomposing it into:
\begin{align}\label{eq:H}
    \vb{H} \approx \vb{QA},
\end{align}
where
\begin{align}
    \vb{Q} = \begin{pmatrix}
        \mid & \mid & & \mid \\
        \vb{q}_{1}, & \vb{q}_{2}, & \cdots, & \vb{q}_{K}\\
        \mid & \mid & & \mid \\
    \end{pmatrix},
    \qq{}
    \vb{A} = \begin{pmatrix}
        \text{---} & \vb{a}_{1} & \text{---} \\
        \text{---} & \vb{a}_{2} & \text{---} \\
        & \vdots &  \\
        \text{---} & \vb{a}_{K} & \text{---} \\
    \end{pmatrix},
\end{align}
where $\vb{Q}\in \mathbb{R}^{M^2\times K}$ and $\vb{A}\in \mathbb{R}^{K\times M^2}$ contain the $K$ flattened eigenPSFs, $\vb{q}_k$, and eigencoefficients, $\vb{a}_k$, respectively.
An efficient computation of $\vb{g}\approx \vb{QAf}$ is as follows (Fig.~\ref{fig:eigencwd_schematic}):
\begin{enumerate}
    \item Element-wise multiplication of the object, $\vb{f}$, with the $k$th eigencoefficients, $\vb{a}_k$.
    \item Convolve the result with the $k$th eigenPSF, $\vb{q}_k$, which yields an eigenImage.
    \item Sum all $K$ eigenImages to obtain the spatially-varying blurred object.
\end{enumerate}

In the original implementation of CWD-SVD \cite{Sroubek2016-ki}, Sroubek used the SVD approach to obtain the basis PSFs to populate the matrix $\vb{Q}$, followed by computing $\vb{A} = \qty(\vb{Q}^T\vb{Q})^{-1}\vb{Q}^T\vb{H}$ to obtain the associated weights for Eq.~\eqref{eq:H}.
However, the purpose of PSF decomposition described in the previous section is to avoid the need to construct the large matrix $\vb{H}$ (size $M^2\times M^2$), which renders Sroubek's approach of calculating $\vb{A}$ computationally inefficient since it requires precomputing $\vb{H}$.
Here, we instead use the eigenPSF approach described in Sec.~\ref{sec:eigenpsf} to directly compute the eigencoefficients to populate $\vb{A}$ through interpolation.
Therefore, our modified eigenCWD algorithm is capable of scaling to much larger $M$ as the necessary matrices to model image formation ($\vb{Q}$ and $\vb{A}$) only have sizes of $M^2\times K$, where $K\ll M^2$.

To perform spatially-varying deconvolution, eigenCWD seeks a solution for the object, $\vb{f}$, which minimizes the following objective function:
\begin{align}\label{eq:loss}
    \min_{\vb{f}} \frac{\mu}{2} \norm{\vb{QAf} - \vb{g}}^2_2 + \alpha \norm{\nabla\vb{f}}_1,
\end{align}
where $\nabla$ is the gradient operator and $\alpha$ is a hyperparameter.
The second term in Eq.~\eqref{eq:loss} is the total variation (TV) operator \cite{Rudin1992-ij} which regularizes the problem.
We employ the same strategy of using the ADMM method to solve this minimization problem -- details may be found in the original paper by Sroubek \cite{Sroubek2016-ki}.

\section{Results and discussions}

In the following subsections, we demonstrate the efficacy of both eigenPSF in approximating the forward blurring model, as well as eigenCWD in recovering the deblurred object for a metalens imparting a hyperbolic phase profile in the incident light.
These lenses are widely used in the community as they provide large focusing efficiencies at normal incidence, but are also known to suffer from severe off-axis aberrations, limiting their use for imaging applications. Their phase profile has the following form:
\begin{align}
    \phi\qty(r) = \begin{cases}
        \frac{2\pi}{\lambda}\qty(f - \sqrt{r^2 + f^2}), & r<\frac{D}{2} \\
        0, &\text{otherwise,}
    \end{cases}
\end{align}
where $r=\sqrt{x^2+y^2}$ is the radial distance from the optical axis, $\lambda$ is the wavelength, and $f$ and $D$ are the focal length and diameter of the lens.
Here, we simulated a $D =$ \SI{200}{\um} lens on a $601 \times 601$ grid with pixel pitch of \SI{400}{\nm}, with $\lambda =$ \SI{740}{\nm} and $f=$ \SI{173}{\um}.
A square object with a length of $L=\SI{1.25}{\m}$ is placed $s=\SI{2}{\m}$ away from the hyperbolic lens, corresponding to an angular field of view of \SI{46.9}{\degree}.
The blurred image is formed in the focal plane of the lens.
We crop the blurred image to a size of $375 \times 375$ pixels for deblurring with eigenCWD.

The PSF, $p$ corresponding to a point-emitter at the object coordinate $\qty(u,v)$ is numerically computed using
\begin{align}
    p(u,v,x,y) = \mathcal{P}_f\qty{e^{i\frac{2\pi}{\lambda}\qty[x\sin{\theta_x\qty(u)} + y\sin{\theta_y\qty(v)}]}\phi(x,y)},
\end{align}
where $\theta_x = \arctan{\frac{u}{s}}$ and $\theta_y = \arctan{\frac{v}{s}}$ are the horizontal and vertical angles the point emitter makes with respect to the optical axis, and $\mathcal{P}_f$ is the angular spectrum propagator \cite{Yu2012-ff} which propagates the wave from the lens to the focal plane.

Both the structural similarity index measure (SSIM) \cite{Wang2004-qe} and the peak signal-to-noise ratio (PSNR) metrics are used to quantify the similarity between two images.

\subsection{Image formation with eigenPSF}
\begin{figure}[htbp]
    \centering
    \includegraphics[width=0.8\textwidth]{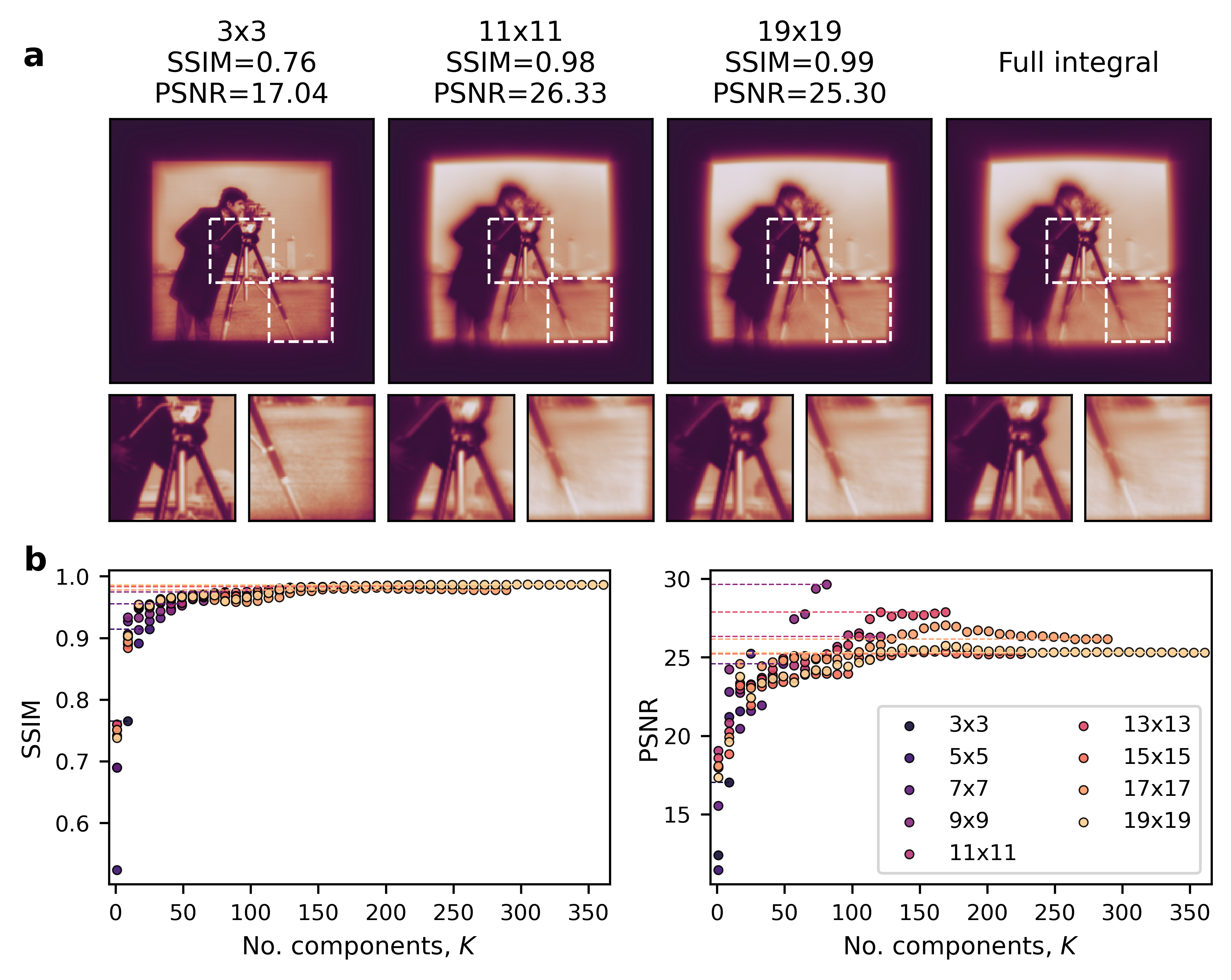}
    \caption{
    Approximation of forward blurring model with eigenPSF method. 
    (a) The computed images using $3\times 3$, $11\times 11$ and $19\times 19$ sampled PSFs, compared against evaluating the explicit integral in Eq.~\eqref{eq:rs}. 
    The SSIM and PSNR are computed between the eigenPSF simulated images against that of the explicit integral. 
    (b) The SSIM and PSNR of the eigenPSF blurred images with varying numbers of sampled $n \times n$ PSFs and eigenPSF components.}
    \label{fig:image}
\end{figure}

We first compare the accuracy of the eigenPSF method (Eq.~\eqref{eq:eigenconvolution}) against evaluating the full integral in Eq.~\eqref{eq:rs} for computing the spatially blurred image.
Fig.~\ref{fig:image} depicts the eigenPSF simulated images using an increasing number of sampled PSFs on an evenly spaced grid with sizes ranging from $3 \times 3$ to $19 \times 19$ (see Supplementary Information for images for all grid sizes).
In general, we observe that increasing the number of sampled PSFs results in improved SSIM but at the expense of increasing computational time.
Furthermore, the SSIM plot in Fig.~\figref[b]{fig:image} also shows that truncating the number of eigenPSFs (or components), for a larger number of sampled PSFs, has negligible effects in approximating the spatially-varying blur.
For example, while the $19 \times 19$ case has a total number of 361 eigenPSFs, keeping only up to $\sim 150$ is sufficient to achieve the highest SSIM possible with this approach.

\begin{figure}[htbp]
    \centering
    \includegraphics[width=0.5\textwidth]{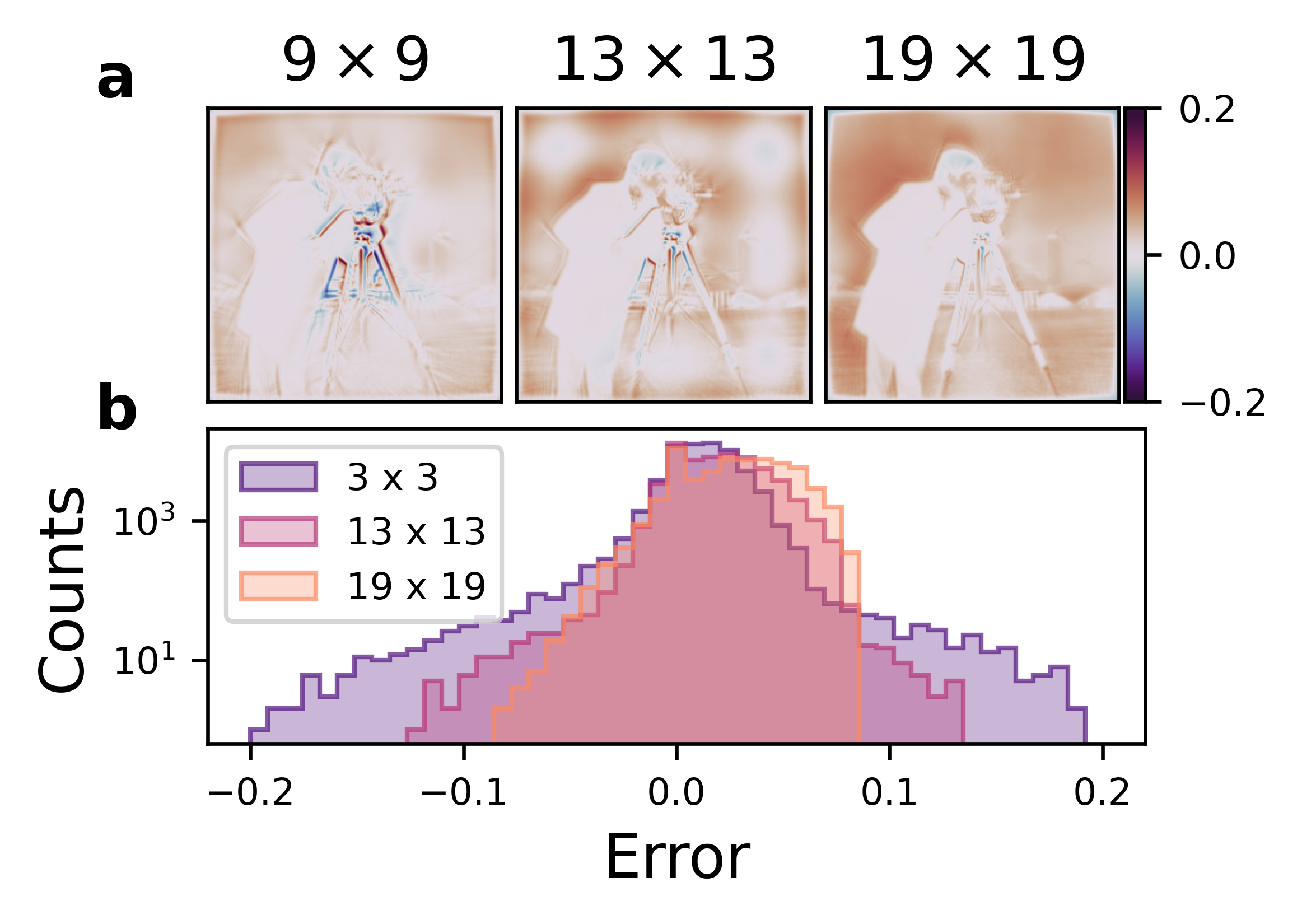}
    \caption{
    Pixel-wise difference between the computed blurred images using the eigenPSF method against evaluating the explicit integral. 
    At a lower number of sampled PSFs, larger errors are concentrated at the center of the image. 
    Increasing the number of sampled PSFs distributes this error more evenly across the whole image.
    The lowering of pixel-wise error increases SSIM, but the distribution of the error across more pixels in the image causes a decrease in PSNR.}
    \label{fig:error}
\end{figure}

However, the PSNRs in Fig.~\figref[b]{fig:image} do not reflect the trend observed in the SSIM plots.
To see why, we refer to Fig.~\ref{fig:error} which depicts the pixel-wise error between the images computed using the eigenPSF formalism (Eq.~\eqref{eq:eigenconvolution}) and the full integral approach (Eq.~\eqref{eq:rs}) for increasing number of sampled PSFs.
Here, we see that the error contributions gradually spread from the center to the rest of the image with increasing PSF samples.
Although the actual per-pixel errors become smaller, this spreading results in the overall inflation of pixel values, leading to a lower reported PSNR for a larger number of sampled PSFs.
This pixel value inflation may be because the interpolation method used to obtain the eigencoefficients, $\vb{A}$, does not guarantee that the effective PSFs at each pixel location (Eq.~\eqref{eq:sumeigenpsf}) are normalized \cite{Lauer2002-au}.
This sometimes leads to hot-spotting artifacts most evident in the $17 \times 17$ blurred image in the Supplementary Information.
As a result, the total pixel value is not conserved when using eigenPSFs to calculate the spatially-varying blur.
This leads to a lower reported PSNR, whereas the SSIM would remain relatively unaffected.

\subsection{Reconstruction with eigenCWD}
\begin{figure}[htbp]
    \centering
    \includegraphics[width=0.8\textwidth]{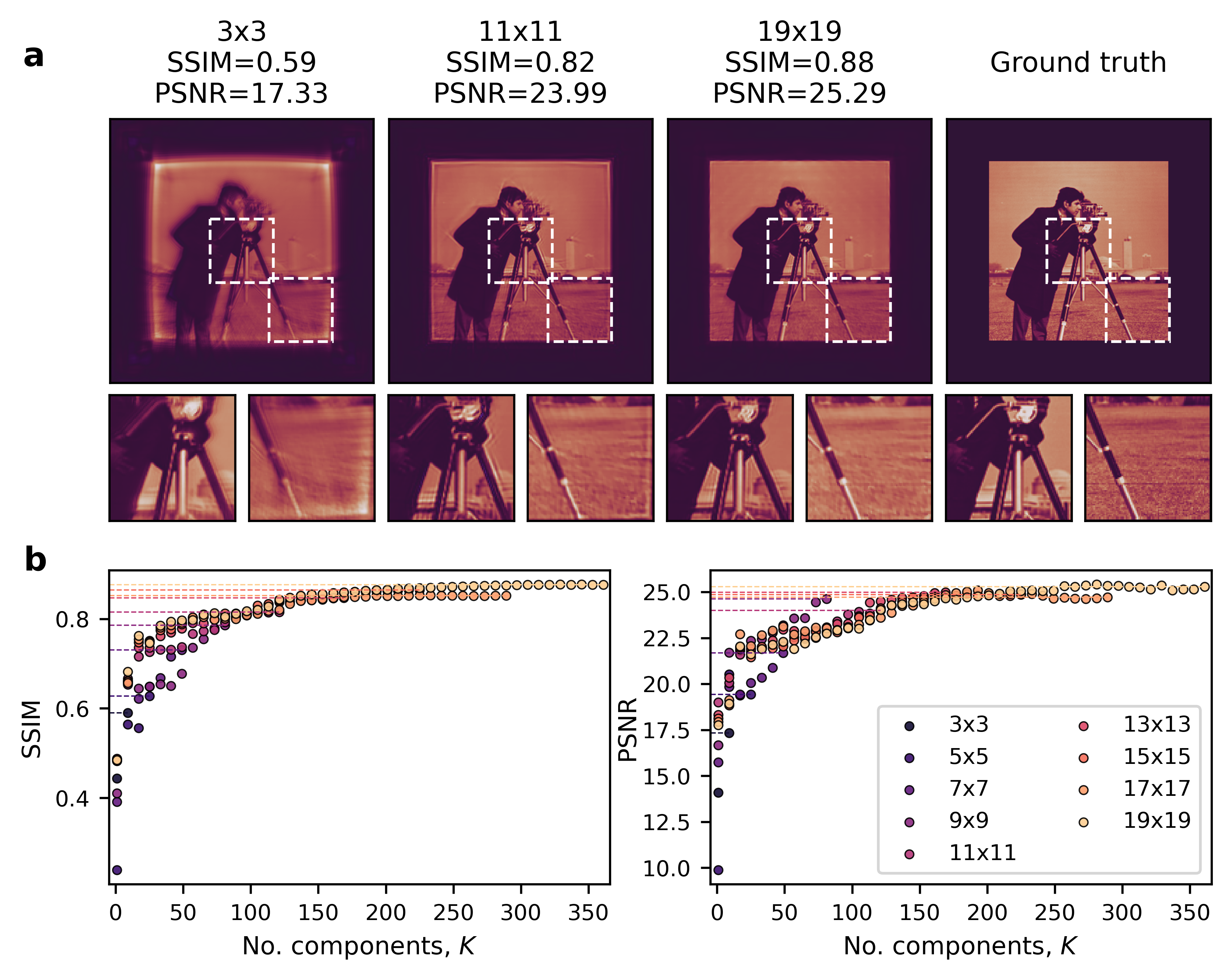}
    \caption{
    spatially-varying deconvolution via eigenCWD.
    The image to be deconvolved is simulated using the explicit integral in Eq.~\eqref{eq:rs}.
    (a) The reconstructed object using $3\times 3$, $11\times 11$ and $19\times 19$ sampled PSFs, compared against the ground truth. 
    The SSIM and PSNR are computed between the eigenCWD reconstructions and the ground truth. 
    (b) The SSIM and PSNR of the reconstructions using the eigenCWD method with varying numbers of sampled $n \times n$ PSFs and eigenPSF components.}
    \label{fig:recon}
\end{figure}

Here, we apply the eigenCWD algorithm described in Sec.~\ref{sec:eigencwd} to perform spatially-varying deconvolution on an image simulated using the full integral approach (rightmost image in Fig.~\figref[a]{fig:image}) to recover the object.
EigenCWD was implemented on a single NVIDIA L40 GPUs, and the runtimes ranged from \SI{4}{\s} ($3 \times 3$ sampled PSFs) to \SI{160}{\s} ($19 \times 19$ sampled PSFs) for 4000 iterations with $\mu = 10^5$ and $\alpha = 1$.

Fig.~\ref{fig:recon} shows the reconstruction obtained with eigenCWD using an increasing number of sampled PSFs (see Supplementary Information for reconstructions for all grid sizes).
Again, we observe improvements in reconstruction fidelity with increasing number of sampled PSFs.
Fig.~\figref[b]{fig:recon} also demonstrates that with a sufficient number of sampled PSFs, it is possible to truncate the number of eigenPSF components whilst maintaining reconstruction quality (for example, keeping $K \approx 200$ for $19 \times 19$ sampled PSFs is sufficient).

However, sampling $N_1$ PSFs and using all $K_1 = N_1$ components is not the same as sampling a larger number of $N_2 > N_1$ PSFs but keeping the first $K_1$ components.
For example, Fig.~\figref[b]{fig:recon} shows that keeping all components of $17 \times 17$ sampled PSFs yields a reconstruction with lower SSIM than keeping the first $17^2 = 289$ components of the $19\times 19$ variant.
Sampling more PSFs results in improved interpolation for the weights, $\vb{A}$, and therefore the $19 \times 19$ sampled PSFs better describe the spatially-varying blur due to the lens compared to $17 \times 17$.
On the other hand, truncating the number of components keeps the first $K$ eigenPSFs, removing the components with minimal loss in information.
Therefore, sampling a larger number of PSFs followed by truncating the number of components would allow for faster reconstruction with negligible loss in quality.

\begin{figure}[htbp]
    \centering
    \includegraphics[width=\textwidth]{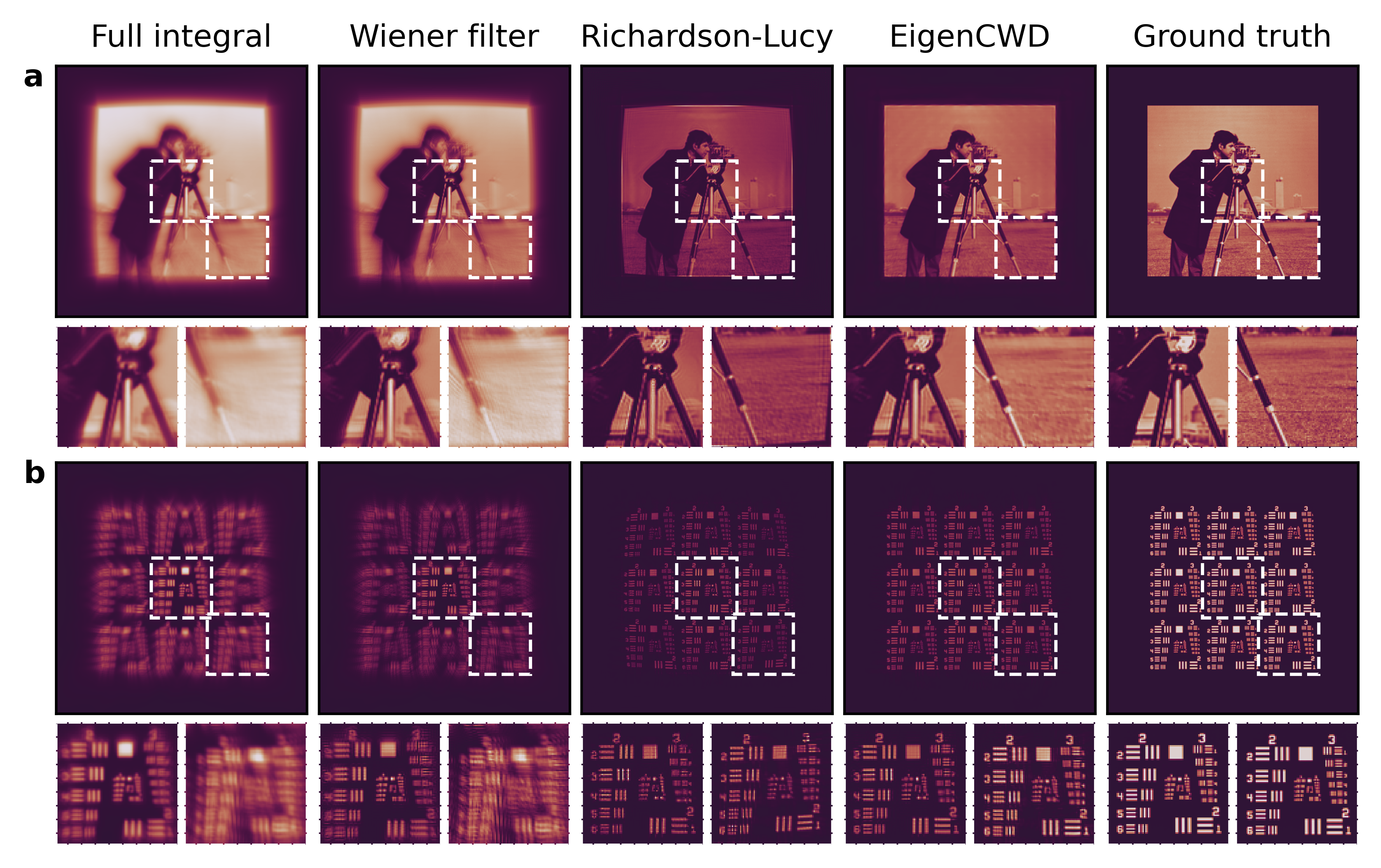}
    \caption{
    Comparison of image deblurring using the Wiener filter and our proposed eigenCWD for (a) cameraman and (b) USAF1951.
    }
    \label{fig:hyper}
\end{figure}

We also compare the eigenCWD algorithm against the commonly used Wiener filter \cite{Wiener2019-je} in Fig.~\ref{fig:hyper}.
For the hyperbolic lens which suffers from severe coma aberration, it is evident that the Wiener filter fails to correct for both the barrel distortion and off-center aberrations.
In contrast, eigenCWD effectively reconstructs the object by accounting for the varying coma aberration without requiring explicit knowledge of the blur kernels at every pixel location.

We also performed a similar comparison for both spherical and parabolic lens profiles (see Supplementary Information).
While images formed with both these lens profiles do not exhibit severe coma aberrations like the hyperbolic lens, they result in a stronger barrel distortion, for which eigenCWD implicitly corrects as well.

\section{Conclusion}
In this work, we have developed and applied the eigenCWD, a spatially-varying deconvolution algorithm based on the eigenPSF decomposition method.
With it, we demonstrated its potential to correct spatially-varying blur, as well as geometric distortions, commonly found when imaging with single-metalens cameras.
We also demonstrated the feasibility of minimizing both speed and memory overheads by truncating the least significant eigenPSF components, with negligible impact on reconstruction quality, thus paving the way for its use in the post-processing of images captured by these types of systems.

We should point out a limitation of eigenCWD, which is the need for the imaged object to be finitely supported.
A nonzero background around the object of interest can cause ringing artifacts to contaminate the reconstructed object, which has been observed when applying eigenCWD to experimental data \cite{Baranikov2023-hz}.
This is because the imposed periodicity of the discrete Fourier transform at the edges causes pixels from the neighboring periodic copy to be deconvolved into the main image.
These ringing artifacts, however, may be suppressed by incorporating edge-smoothing strategies into eigenCWD \cite{Renting_Liu2008-xd}, which could be subject to further studies in the future.

\paragraph{Funding}    
The authors acknowledge support from the Agency for Science, Technology and Research of Singapore under the A*STAR Graduate Scholarship, the AME Programmatic Grant, Singapore, under Grant A18A7b0058, and the support of the Early Career
Research Award from the National University of Singapore (NUS).

\paragraph{Acknowledgements}    
The authors would like to acknowledge insightful conversations with Anton V. Baranikov.

\paragraph{Data availability}
The codes for the eigenPSF method and the eigenCWD algorithm are available at \url{https://doi.org/10.5281/zenodo.13329612}.
    
\paragraph{Supplemental document}
See Supplement 1 for supporting content.



\bibliographystyle{unsrt}
\bibliography{sample}

\newpage

\appendix

\begin{center}
    \textbf{{\huge Supplementary Information}}
\end{center}
\numberwithin{equation}{section}
\setcounter{equation}{0}

\section{EigenPSF forward model and eigenCWD reconstruction for hyperbolic lens for all gridsizes}
\begin{figure}[!h]
    \centering
    \includegraphics[width=\textwidth]{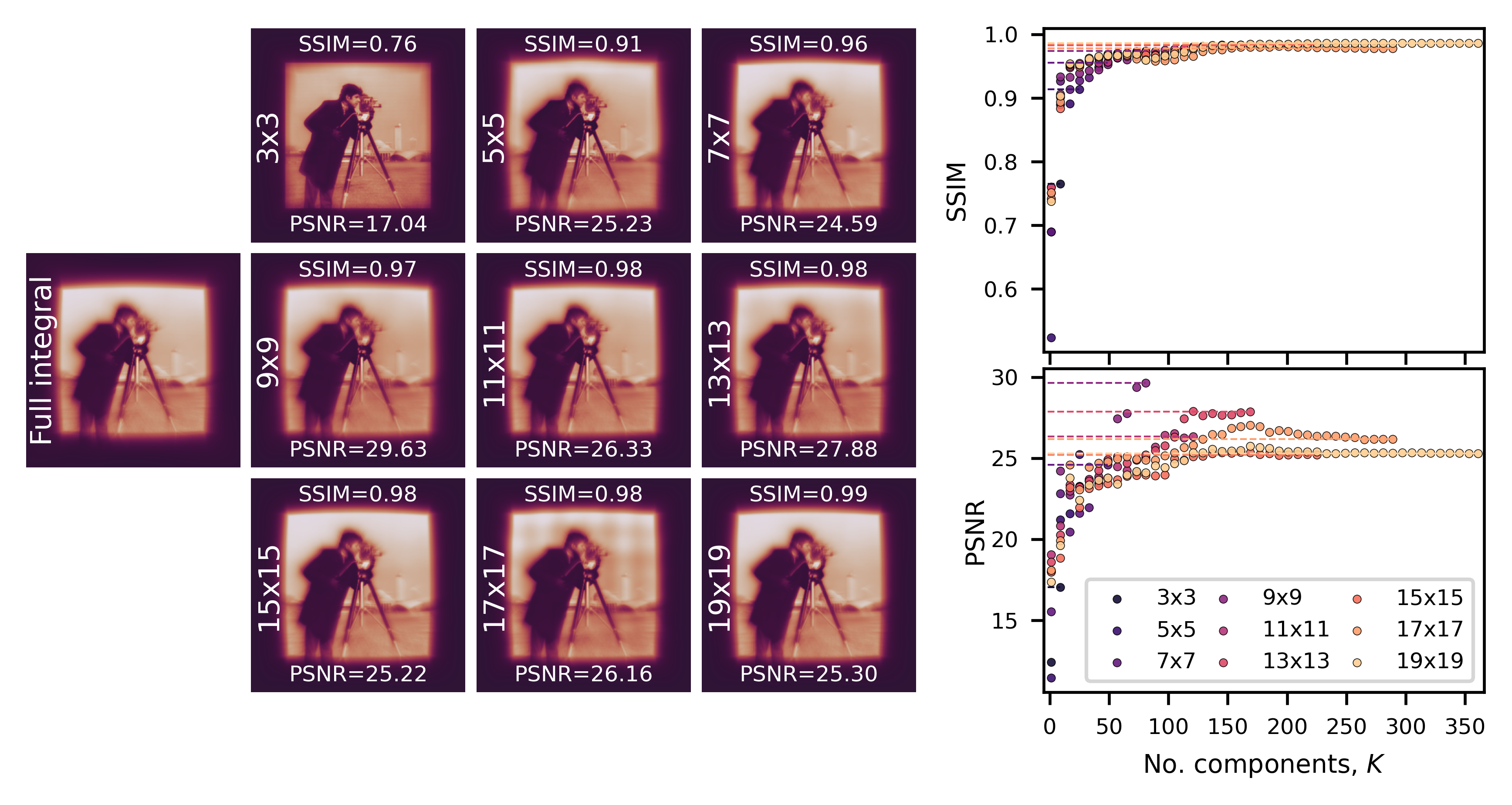}
    \caption{EigenPSF forward model for all sampled PSFs gridsizes.}
    \label{fig:allsizeblur}
\end{figure}
\begin{figure}[htbp]
    \centering
    \includegraphics[width=\textwidth]{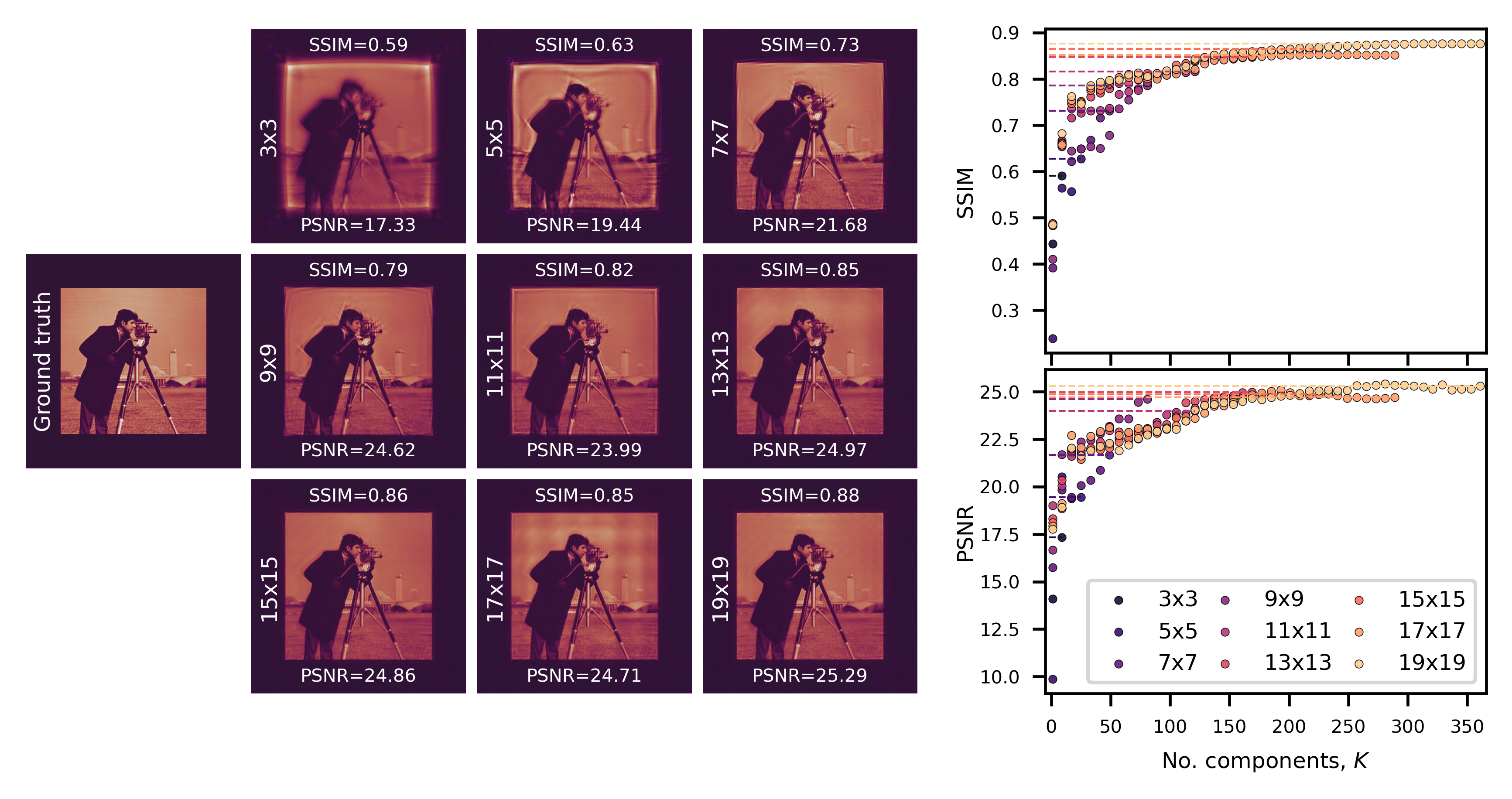}
    \caption{EigenCWD reconstruction for all sampled PSFs gridsizes.}
    \label{fig:allsizerecon}
\end{figure}

\newpage

\section{Time taken for eigenPSF forward model and eigenCWD reconstruction for varying number of components}
\begin{figure}[!h]
    \centering
    \includegraphics[width=\linewidth]{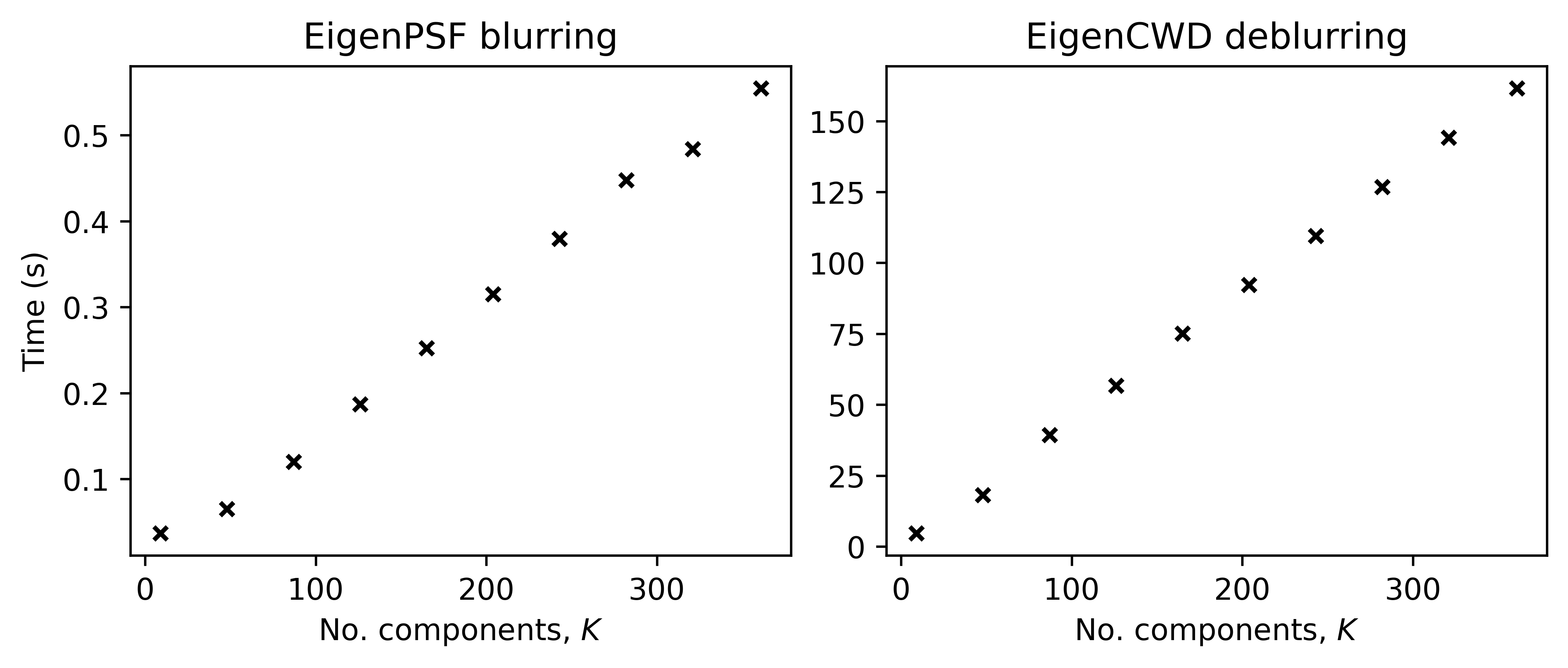}
    \caption{Time taken to perform (left) eigenPSF forward model blurring and (right) eigenCWD reconstruction on a single NVIDIA L40 GPU.
    The timings here assume that the eigenPSFs and the eigencoefficients have been precalculated already.}
    \label{fig:enter-label}
\end{figure}

\section{EigenCWD reconstruction for various lens profiles}
\begin{figure}[!h]
    \centering
    \includegraphics[width=\textwidth]{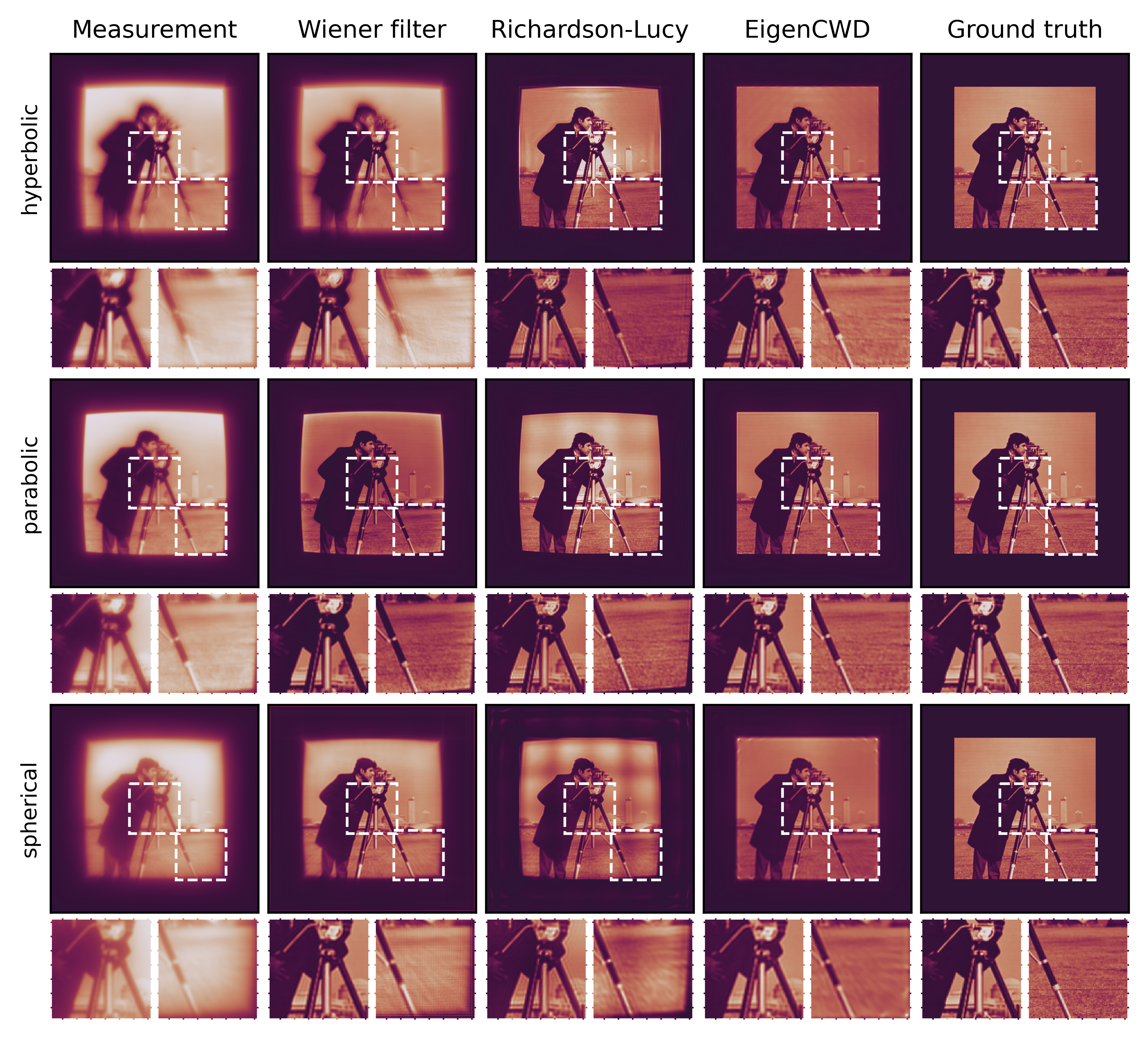}
    \caption{
    Reconstruction of cameraman for hyperbolic, parabolic and spherical lenses. 
    All parameters of the lenses here are the same as that of the hyperbolic as described in the main text.
    However, the measurements for the parabolic and spherical lenses are obtained at a distance of \SI{163}{\um} from the lens instead (as opposed to \SI{173}{\um} for hyperbolic), corresponding to the smallest PSF on the optical axis.}
    \label{fig:allreconcamera}
\end{figure}
\begin{figure}[!h]
    \centering
    \includegraphics[width=\textwidth]{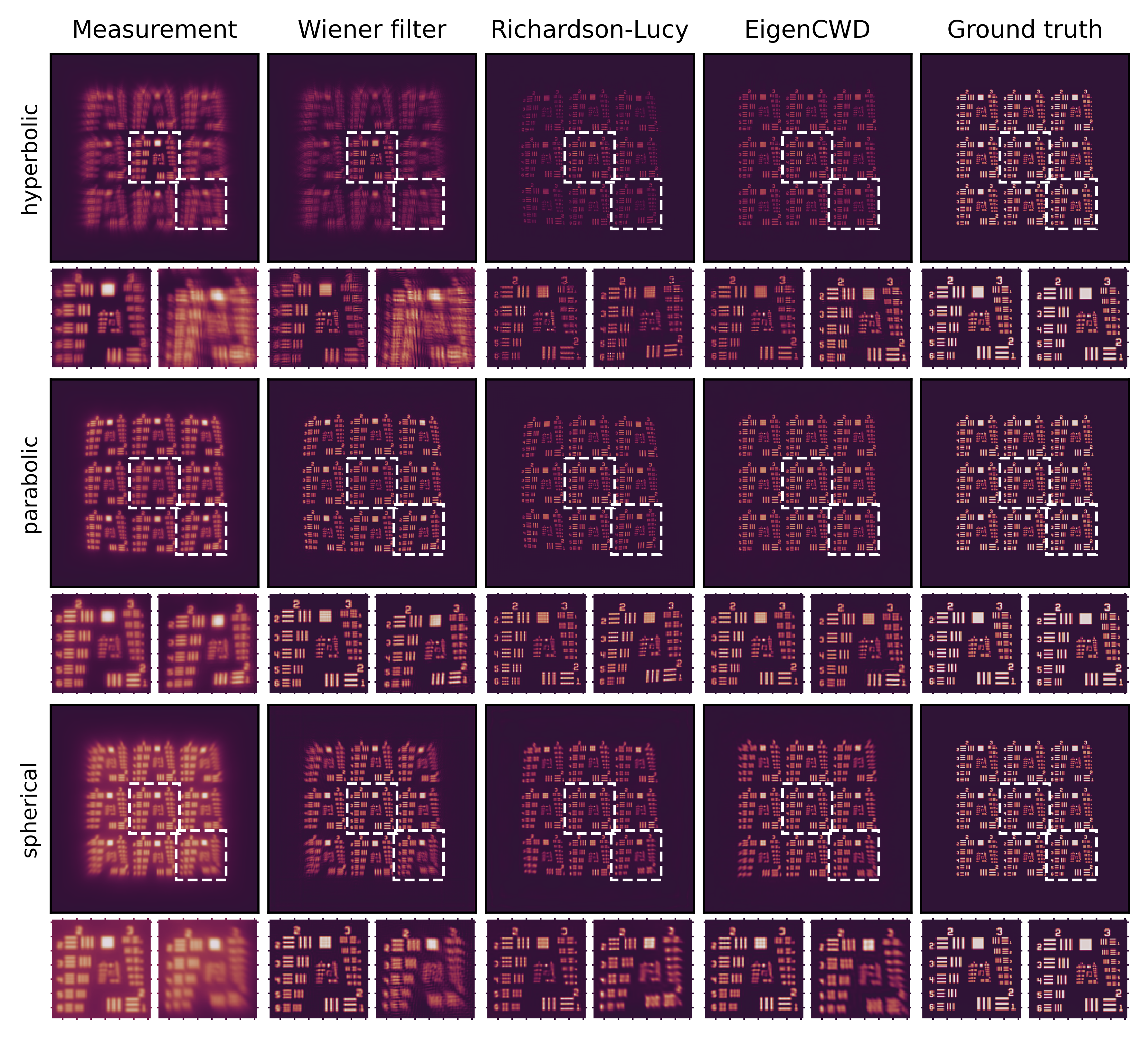}
    \caption{
    Reconstruction of USAF1951 for hyperbolic, parabolic and spherical lenses. 
    All parameters of the lenses here are the same as that of the hyperbolic as described in the main text.
    However, the measurements for the parabolic and spherical lenses are obtained at a distance of \SI{163}{\um} from the lens instead (as opposed to \SI{173}{\um} for hyperbolic), corresponding to the smallest PSF on the optical axis.}
    \label{fig:allreconusaf}
\end{figure}
Fig.~\ref{fig:allreconcamera} and~\ref{fig:allreconusaf} compares the reconstruction between EigenCWD and Wiener filtering for three lens profiles: hyperbolic, parabolic and spherical.
The phase profiles are defined as follows:
\begin{align}
    \text{Hyperbolic: }& \phi\qty(r) = \begin{cases}
        \frac{2\pi}{\lambda}\qty(f - \sqrt{r^2 + f^2}), & r<\frac{D}{2} \\
        0, &\text{otherwise,}
    \end{cases} \\
    \text{Parabolic: }& \phi\qty(r) = \begin{cases}
        -\frac{\pi r^2}{\lambda f}, & r<\frac{D}{2} \\
        0, &\text{otherwise,}
    \end{cases}\\
    \text{Spherical: }& \phi\qty(r) = \begin{cases}
        \frac{2\pi}{\lambda}\qty(\sqrt{\qty|f^2-r^2|} - f), & r<\frac{D}{2} \\
        0, &\text{otherwise,}
    \end{cases}
\end{align}

EigenCWD and Richardson-Lucy significantly improves Wiener filtering for the hyperbolic lens, due to the severe coma aberration inherent in this lens profile that the Wiener filter cannot correct for.
On the other hand, both parabolic and spherical lens profiles exhibit much less variation in the PSFs at this field of view, therefore resulting in much more similar results across Wiener filter, Richardson-Lucy and eigenCWD.
However, the barrel distortion remains in both the Wiener filter and Lucy-Richardson results, whereas eigenCWD can also correct for such geometric distortions.
In addition, reconstruction artifacts such as edge-ringing and hot-spotting are evident in the Richardson-Lucy reconstruction of the cameraman.

\end{document}